# Space Program Language (SPL/SQL) for the Relational Approach of the Spatial Databases[*]

I. Vega-Páez and Feliú D. Sagols T.[1]

e-mail:
vegapaez@mvax1.red.cinvestav.mx
fsagols@mvax1.red.cinvestav.mx

```
Department of Electric Engineering
       Section to Computer
Center of Investigation and Advanced Studies of the
      Polytechnic National Institute,
             México, D.F.
```

## Abstract

In this project we are presenting a grammar which unify the design and development of spatial databases. In order to make it, we combine nominal and spatial information, the former is represented by the relational model and latter by a modification of the same model. The modification lets to represent spatial data structures (as Quadtrees, Octrees, etc.) in a integrated way. This grammar is important because with it we can create tools to build systems that combine spatial-nominal characteristics such as Geographical Information Systems (GIS), Hypermedia Systems, Computed Aided Design Systems (CAD), and so on.

## Introduction

Normally based of spatial data are systems where the spatial information is stored, processed and analyzed. Such systems understand between other Geographic Information Systems (GIS's), Systems of Computer Aided Design and Manufacture (CAD/CAM) and most recently Hypermedia Systems. Nevertheless to the crescent diffusion, in majority of databases that require combine nominal and spatial information, the last representation appears forced because this type of databases were developed originally using DBMS without support for Spatial Data, or well did using proper systems for the image processing or computer graphics.

A new proposal of unification, based in the relational model and the properties of some data structure as known lineal quadtrees (Li and Loew, 1987), proposes initially by Wang (Wang, 1990) for classification of remote sensing images, it is modified and formalized for facilitate the representation of spatial information (Vega-Páez and F. Sagols, 1994). This formulation drift of a work prior presented by Samet (Samet et al., 1984). Our approach combines advantages of the relational databases model with hierarchical data structures quadtrees (Samet, 1990a and 1990b) and some other structures (Preparata and Shamos, 1985), proportioning tools for spatial-nominal schemes.

[1] Members of the Section of Computing of the CINVESTAV (Center of Investigation and Advanced Studies of the Polytechnic National Institute, México)
[*] This work was supported in part by Council National of Science and Technology (Consejo Nacional de Ciencia y Tecnología CONACYT).

The spatial databases are characterized by containing representations of spatial entities and relationships of the objects, that integrate to spatial information and nonspatial information. To model such systems require of data schemes capable representing as spatial entities as the nominal information of such manner that support operations of manipulation and consultation of four types :

- *Topology Operations:* They are related operations with aspects Topologies such as adjacentence between geometric elements, frontiers, connectiveity, and so on. Examples of these operations we have upon asking in a map by the neighboring countries to a given country or by the description of the frontier of a specific country.
- *Set Operations:* They are classical operations of the set theory applied to geometric elements to see as sets of points. An example of this when upon modeling a table in a solid computer aided design system, we are utilizing the union of a set of parallelepiped.
- *Metric Operations:* These operations base on the notion of distance between two points (normally use the Euclidean distance). An example of this is the how meter of construction in a architectural design system with ends to budget you and of control of working.
- *Spatio-nominal Operations:* They are operations that involve so the management of nominal properties in geometric elements as the management of geometric properties in nominal elements. For example, in an urban system does an operation of this type upon asking so sign in a map the intersection of two whose avenues names have specified previously.

In short, the combination of hierarchical structure with nominal entities in the relational model, in conjunction with the prior operations, constitutes the base to design Spatial DataBases Management Systems.

## Methodology

To realize our proposal of solution to the prior problem, was proposed the following plan:
- Analysis of problematic in Representation of spatial information. Where define the spatial basic elements, problematic of representation so in the relational model as in hierarchical structures and the combination, for the bidimensional case.
- Analysis of lineal quadtree and the properties for the bidimensional case.
- Propose of relational approach of the lineal quadtree to represent structure spatial basic and the operations with the structure lineal quadtree. For it which define schemes of relation that represent the basic elements point, line, area, and so on. And the operations that can realize over they with expressions type of Structured Query Language (SQL) with sole to add operators that involve lineal quadtree codes and quadtree operators (Vega-Páez I. and F. Sagols, 1994).
- Realize a tool that has an extent of SQL that includes to the quadtree operators and the schemes of the basic elements. An alternative is to realize in a frame client-server technology, that counts with a basic nucleus that has our extent in the server and a scheme of interaction with the customer in a language guest of manner imbibed SQL or in the form pure programming of language guest with called to routines to the style of a API (Application Programming Interface). Other alternative is in format imbibed SQL as language of programming with varied types of support to commercial databases.

Have planned to realize three practical applications that use prior tool demonstrating the power of extent for the spatial systems, that in our case is toward Hypermedia Systems, Geographic Information Systems and Geometric Modeling Systems.

## Results

Until the moment has for the bidimensional case, of the which can obtain the three-dimensional case motive of a back report. In the scope of this paper we are presenting a extent of SQL that includes the quadtree operators, with this extent implant spatial operations, for major information see the proposal of Unification of the Spatial DataBases (Vega-Páez I. and F.Sagols, 1994). This grammar is an extent to called Programming Language for SQL (PL/SQL) proposed by the norms ANSI X3.135-192 (SQL/2, 1992) and ANSI X3H2 1992-109 (SQL/3, 1992), in the year of 1992 (Date and Darwen, 1994). This norms are partially included in the lasts versions of commercial DBMS with ORACLE, INFORMIX, SYBASE, between others.

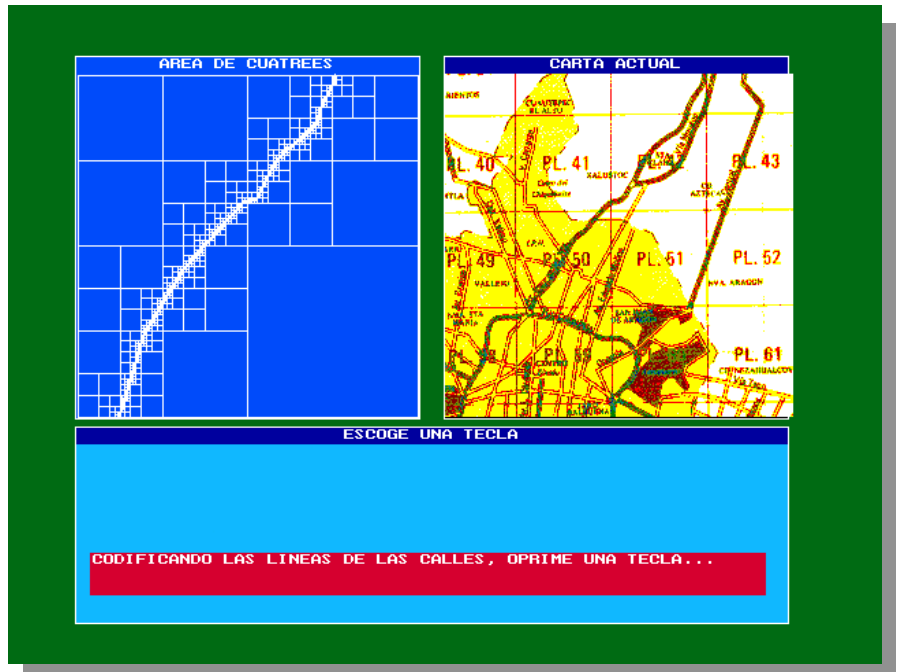

**Fig. 1. The Insurgentes street and coding in Quadtree for Database**

Other result is an editor of maps, that permits to code geometric entities to be stored into database, as was exemplified in the figures [1] and [2].

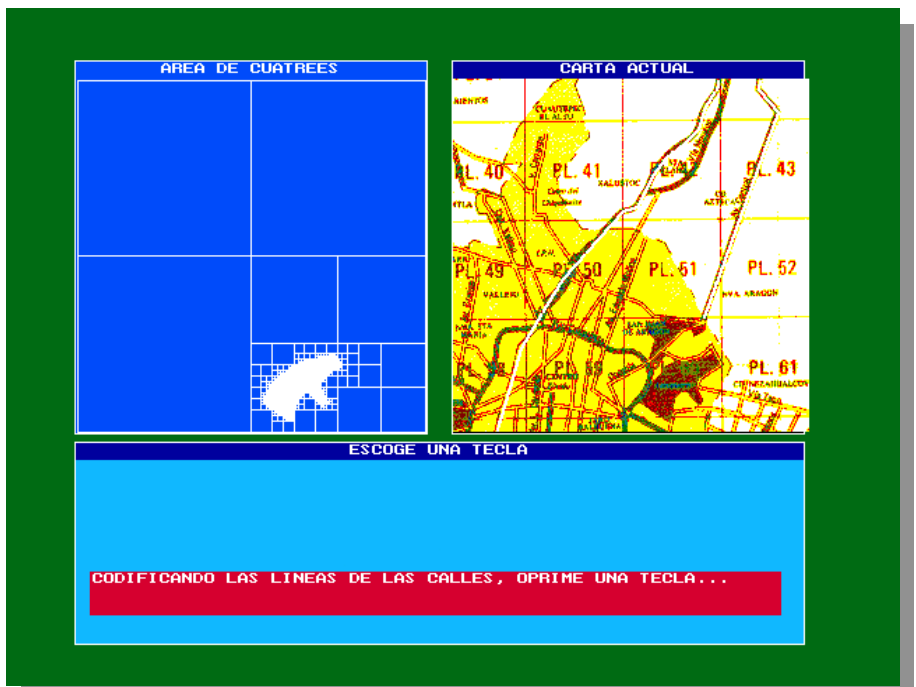

**Fig. 2. Closed area (Airport of México City) and Coding in Quadtree for Database**

SPL/SQL

The version is the syntax in BNF format of Spatial Program Language SQL, not contained ambiguities and includes the quadtree operators exposed in Relational Approach of the Spatial DataBases (Vega-Páez I. and F. Sagols, 1994).

For the development of the software has the project of the tool with client-server technology for SPL/SQL. Also this writing the theory for the three-dimensional case and the fundamentals of digital neighborhood as part of the call *Digital Topology*.

## Examples for SPL/SQL

The examples illustrate the use of Language to implant the spatial operations.
1). Find the intersection point of line A and line B, see figure [3].

```
PROCEDURE INETER_A_B ( SQLSTATE,
                :line_a_param CHAR(8), :line_b_param CHAR(8) );
SELECT CODE
     FROM LINES
     WHERE LINE = :line_a_param
   INTERSECT
    SELECT CODE
     FROM LINES
     WHERE LINE = :line_b_param;
```

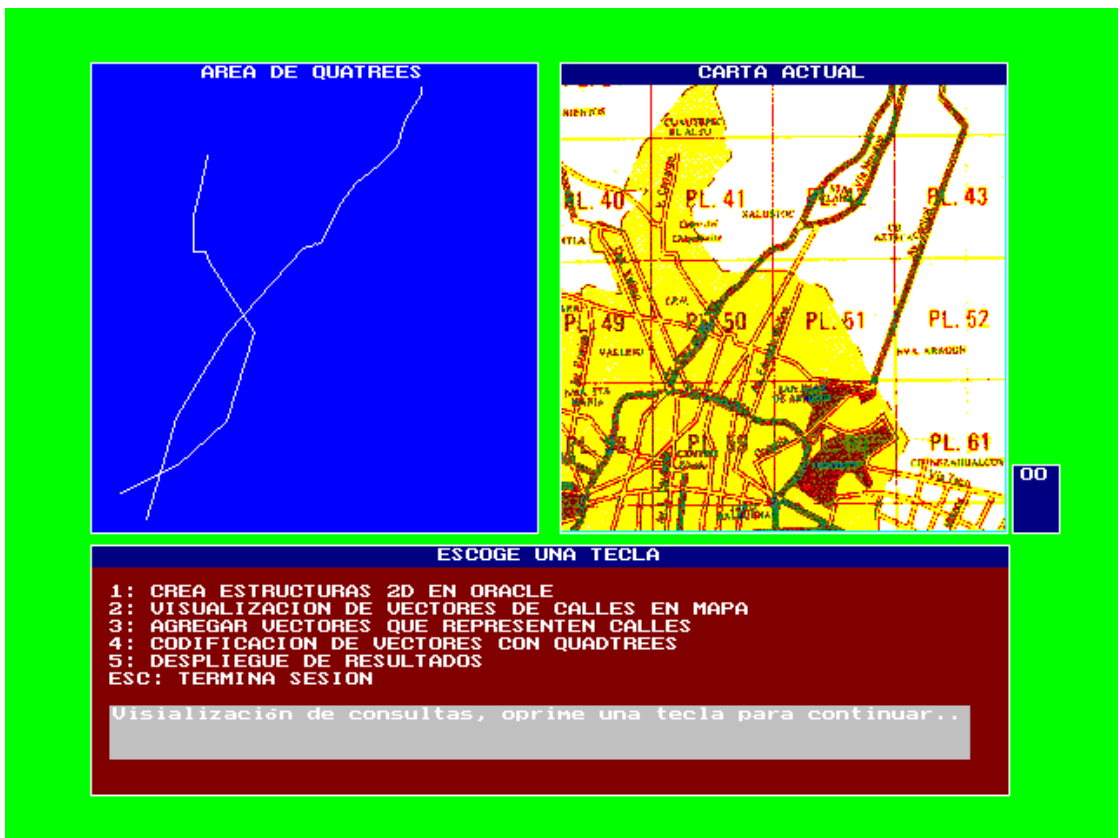

**Fig. 3. Example of intersection two streets (poli-lines) in México, city**

2). Find all lines crossing point A.

PROCEDURE ALL_LINES_A ( SQLSTATE, :point_a_param CODE );

```
SELECT LINE
       FROM LINES
MINUS
(SELECT LINE
       FROM  (SELECT *
                     FROM   (SELECT LINE FROM LINES),
                            (SELECT CODIGO FROM POINTS
                                   WHERE POINT=:ponit_a_param)
              )
       MINUS
       LINES)
```

## Conclusion

This technique of representation provided information in multiple resolution to describe together spatial characteristics with nominal information in desired schemes, proportioning a spatial powerful and flexible analysis. An additional advantage of this approach is that the principle is enough clear and of very easy to implant. The only extent to the Language of Programming/SQL is a group of quadtree operators that involve arithmetical operations over informing. In the context of the utilization of DBMS's the quadtree operators can be coded as routines in a guest language (or imbibed) other alternative is in the commands language of the DBMS systems. This advantage does that our approach is practical do use of existing DBMS.